\documentclass[slac_one]{revtex4}
\usepackage{graphicx}
\usepackage{fancyhdr}
\usepackage{d0_style}
\begin{document}
\title{Searching for the elusive graviton}
\author{Edgar Carrera for the \D0 Collaboration}
\affiliation{Florida State University, Tallahassee, FL 32306, USA}

\begin{abstract}
We present a search for large extra dimensions in the single photon plus
missing transverse energy channel (Kaluza-Klein graviton production)
performed using $2.7\ {\rm fb^{-1}}$ of data
collected by the \D0 experiment at the Fermilab Tevatron collider.  At
$95\%$ C.L., we set
limits on the fundamental Planck scale $M_{D}$ from $970\ \text{GeV}$ to
$816\ \text{GeV}$ for two to eight extra dimensions.
\end{abstract}

\maketitle

\thispagestyle{fancy}


\section{\label{sec:intro} INTRODUCTION} 
The large unexplained difference between the effective Planck scale in the
$4$-dimensional space-time ($M_{Pl}\sim 10^{16}\ \text{TeV}$)
and the electroweak scale ($\sim 1\ \text{TeV} $), 
generally known as the hierarchy problem of the Standard Model (SM), served
as the main motivation for the emergence of theories of large extra
dimensions (LED),
also known as ADD theories~\cite{add}.  They postulate the
presence of $n$ extra spatial dimensions with sizes ($R$) large 
comparared to the
electroweak scale.  
While SM particles are bound to the ordinary $3$-dimensional space ($3$-d
brane), gravitons can penetrate the additional volume in detriment of the
strength of the gravitational field in the $3$-d brane.  The 
hierarchy problem is solved, since the fundamental
Planck scale in the $(4+n)$-dimensional space-time ($M_{D}$), which could
be of the order of the electroweak scale, 
is concealed by the large size of the extra volume: 
$M_{Pl}^2 = 8\pi M_{D}^{n+2}R^{n}$.
The compactification of the extra space forces the gravitational field to
populate only certain energy modes known as Kaluza-Klein (KK) modes.  
Towers of these modes behave like massive, 
noninteracting, stable particles, the $KK$ gravitons ($G_{KK}$), 
whose production
can be inferred in a collider detector by the presence of missing
transverse energy ($\met$).

This review constitutes an update for a previous analysis~\cite{runIIa}, 
where 
we searched for large extra dimensions in the
exclusive (monophoton) channel $q\bar{q}\rightarrow\gamma G_{KK}$ in 
$1\ {\rm fb^{-1}}$ of
data, collected with the \D0~\cite{nim} detector at the Fermilab Tevatron collider.
The present study uses the same analysis techniques on  
$1.7\ {\rm fb^{-1}}$ of additional data.  At the end, we present
the final results as a combination of both analyses ($2.7\
{\rm fb^{-1}}$ of data). Recently,
the CDF collaboration analyzed $2\ {\rm fb^{-1}}$ of data to set 
$95\%$ C.L. lower limits on $M_{D}$, 
from $1080$~\gev~to $900$~\gev~for two to six 
extra dimensions~\cite{cdfrun2}.
Searches for LED in other final states have been performed
by collaborations at the Tevatron~\cite{tevatron,cdfmonojet} 
and the CERN LEP collider~\cite{lep1}.

\section{\label{sec:evt}EVENT SELECTION}

A photon is identified in the detector as a calorimeter 
cluster with at least $90\%$ of 
its energy in the electromagnetic (EM) part.  
The calorimeter isolation variable,
${\cal{I}} = [(E^{\text{tot}}_{0.4}-E^{\text{em}}_{0.2})-\alpha\cdot
l]/E^{\text{em}}_{0.2}$, is required to be less than $0.07$.  In this equation, 
$E^{\text{tot}}_{0.4}$ denotes the total energy deposited
in the calorimeter in a cone of radius
${\cal{R}}=\sqrt{(\Delta\eta)^{2}+(\Delta\phi)^{2}} = 0.4$,
$E^{\text{em}}_{0.2}$
is the EM energy in a cone of radius ${\cal{R}}=0.2$, $l$ is the
instantaneous luminosity, and $\alpha$ is a constant that takes different
values for the central ($|\eta|<1.1$) and end-cap regions ($1.3<|\eta|<4$) 
of the calorimeter. 
The track isolation variable, defined as the scalar sum 
of the transverse momenta of
all tracks that originate from the interaction vertex in an annulus
of $0.05<{\cal{R}}<0.4$ around the cluster, is less than $2\ {\rm GeV}$.
Only EM clusters in the central region of the detector
with both transverse and longitudinal shower shapes consistent
with those of a photon are considered.   The cluster has 
neither an associated track in the central tracking
system, nor a significant density
of hits in the SMT and CFT systems 
consistent with the presence of a track.  Additionally, it is required that
the EM cluster matches an energy deposit in the central preshower (CPS) system.

The {\it photon} sample is created by 
selecting events with only one photon with $p_{T}>90\ {\rm GeV}$, and
at least one reconstructed interaction
vertex consistent with direction of
the photon given by the CPS system. No jets with $p_{T}>15\ {\rm GeV}$ are allowed
in the event.  Jets are reconstructed using the iterative midpoint 
cone algorithm with a cone size of $0.5$.  The missing
transverse energy, which is computed from calorimeter cells with $\eta<4$
and corrected for EM and jet energy scales, is required to be at 
least $70\ \text{GeV}$.  The applied \met~requirement guarantees 
negligible multijet background in 
the final candidate sample, while being
almost fully efficient for signal selection. Events containing muons, 
cosmic ray muons, or tracks with 
$p_{T}>8\ \text{GeV}$ are rejected.

\section{\label{sec:analysis}ANALYSIS}

\begin{figure}[th]
\includegraphics[scale=0.4]{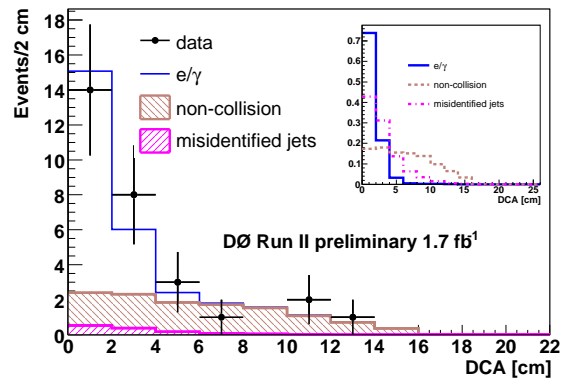}
\caption{\label{fig:fig_1}DCA distribution for the selected events
in $1.7\ {\rm fb^{-1}}$ of 
data (points with statistical uncertainties). The different
histograms represent the estimated background composition from the template
fit to this distribution. The inset figure compares the individual
template shapes.}
\end{figure}

The EM pointing algorithm uses
the fine transverse and
longitudinal segmentation of the calorimeter and the CPS system
to measure the direction of the EM shower. It calculates 
the distance of closest approach 
(DCA) to the $z$ axis (along the beam
line) and predicts the $z$ position of the interaction vertex in the event
independently of the tracker information.
After standardized selection requirements, large backgrounds
to the $\gamma+\met$ signal, which originate from cosmic ray muons
and beam halo particles (non-collision)
depositing energy in the calorimeter, are still present. The discriminating
power of the EM pointing
variables help reduce this background significantly and very efficiently.  
The remaining non-collision events, as well as the contribution from 
$W/Z+\text{jet}$ events where the jet is
misidentified as a photon, are
estimated by performing a linear template 
fit to the data where we exploit the differences in the 
shapes of the DCA distributions (Fig.~\ref{fig:fig_1}). The procedure is described in detail
in~\cite{runIIa}.  

The backgrounds arising from the
process $Z+\gamma \rightarrow \nu\overline{\nu}+\gamma$, which gives the
same signature as the signal, or from $W+\gamma$ 
where the lepton from the $W$ boson decay is not detected, are estimated
from a sample of Monte Carlo (MC) events generated 
with {\sc pythia}~\cite{pythia}, and corrected for luminosity profile
differences with data. Additionally, we apply scale factors to account for
the differences between the efficiency determination in data and
simulation.  $W\rightarrow e\nu$ background, where the
electron is misidentified as a photon, is estimated from data using a
sample of isolated electrons and the measured rate of electron-photon 
mis-identification.

We generate signal events for two to eight
extra dimensions using a modified version of {\sc pythia}~\cite{pythia}.
Table~\ref{tab:summary} gives the final numbers for data and backgrounds.
The main sources of systematic uncertainty are the
uncertainty in the photon
identification efficiency ($5\%$), the uncertainty in
the total integrated luminosity ($6.1\%$), and the uncertainty
in the signal acceptance from the PDFs ($4\%$).  For the standard model
background estimated from MC an uncertainty of $7\%$ in the cross section
is also included.

\begin{table}[t]
\begin{center}
\caption{\label{tab:summary}Data and estimated backgrounds }
\begin{tabular}{|c|c|c|}
\hline

Background&Number of expected events&Number of expected events\\
&($1.7\ {\rm fb^{-1}}$)&(combined analysis, $2.7\ {\rm fb^{-1}}$)\\
\hline
$Z+\gamma \rightarrow \nu\overline{\nu}+\gamma$&$17.4\pm2.2$&$29.5\pm2.5$ \\
$W\rightarrow e\nu$&$4.7\pm1.7$&$8.5\pm1.7$ \\
Non-collision&$3.8\pm1.8$&$6.6\pm2.3$ \\
Misidentified jets&$0.91\pm0.23$&$3.1\pm1.5$ \\
$W+\gamma$&$0.72\pm0.15$&$2.22\pm0.3$ \\
\hline
Total Background&$27.5\pm3.3$&$49.9\pm4.1$ \\
Data&$22$&$51$ \\
\hline
\end{tabular}
\end{center}
\end{table}

\begin{figure}
\includegraphics[scale=0.4]{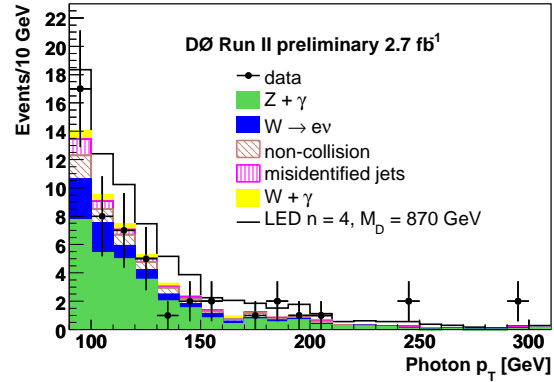}
\caption{\label{fig:fig_2}Photon $p_{T}$ distribution 
for the final candidate events with $2.7\ {\rm fb^{-1}}$ of data after
all the selection requirements.  Data points show statistical
uncertainties. The LED signal is stacked
on top of SM backgrounds.}
\end{figure}

\begin{figure}
\includegraphics[scale=0.4]{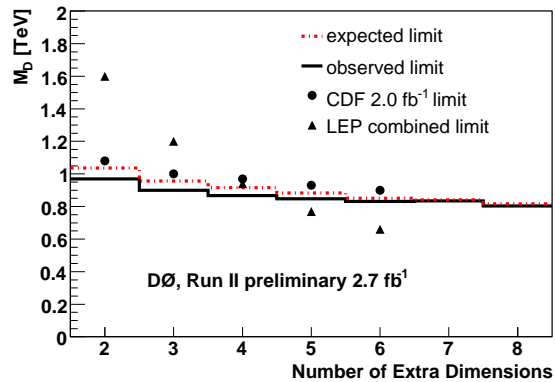}
\caption{\label{fig:fig_3}The 
expected and observed lower limits on $M_{D}$ for
LED in the $\gamma + \met$ final state.  CDF
limits with $2\ \text{fb}^{-1}$~of data (monophoton
channel)~\cite{cdfrun2}, and 
the LEP combined limits~\cite{lep1} are also shown.}
\end{figure}

The total efficiency for the MC signal is $0.38\pm0.04$.
In order to combine this efficiency with the one in the analysis described
in~\cite{runIIa}, we perform a luminosity-weighted
average of the two values and add an extra systematic
uncertainty of $5\%$ due to correlations.  
The combined efficiency is then $0.43\pm0.05$.
Fig.~\ref{fig:fig_2}
shows the photon $p_{T}$
distribution for the combined analysis, with the SM backgrounds stacked on top
of each other.
We employ the modified frequentist approach~\cite{limitc} to
set limits at the $95\%$ C.L. on the production cross section for the signal, assuming 
the leading-order theoretical cross section.
Table \ref{tab:limits} and Fig.~\ref{fig:fig_3} summarize the
limit setting results.


To conclude, we have conducted an update to the analysis described in~\cite{runIIa}
on a search for LED
in the $\gamma+\met$ channel, finding no evidence for
their presence.  The updated limits show significant improvement from
our previous study and are competitive for $n>4$.

\begin{table}[th]
\begin{center}
\caption{\label{tab:limits}Summary of limit calculations.}
\begin{tabular}{|c|c|c|c|c|c|}
\hline
$n$&$1\ {\rm fb^{-1}}$~\cite{runIIa}&$1\ {\rm fb^{-1}}$~\cite{runIIa}&$2.7\ {\rm fb^{-1}}$&$2.7\ {\rm fb^{-1}}$
&CDF $2\ {\rm fb^{-1}}$~\cite{cdfrun2}\\
&observed (expected)&observed (expected)&observed (expected)&observed (expected)&observed\\
&cross section&$M_{D}$ lower&cross section&$M_{D}$ lower&$M_{D}$ lower\\
&limit (fb)&limit (\gev) &limit (fb)&limit (\gev)&limit (\gev)\\

\hline
$2$&$27.6~(23.4)$&$884~(921)$&$19.0~(14.6)$&$970~(1037)$&$1080$\\
$3$&$24.5~(22.7)$&$864~(877)$&$20.1~(14.7)$&$899~(957)$&$1000$\\
$4$&$25.0~(22.8)$&$836~(848)$&$20.1~(14.9)$&$867~(916)$&$970$\\
$5$&$25.0~(24.8)$&$820~(821)$&$19.9~(15.0)$&$848~(883)$&$930$\\
$6$&$25.4~(22.3)$&$797~(810)$&$18.2~(15.2)$&$831~(850)$&$900$\\
$7$&$24.0~(23.1)$&$797~(801)$&$15.9~(14.9)$&$834~(841)$&$--$\\
$8$&$24.2~(21.9)$&$778~(786)$&$17.3~(15.0)$&$804~(816)$&$--$\\

\hline
\end{tabular}
\end{center}
\end{table}

\begin{acknowledgments}

The author wishes to thank Alexey Ferapontov, Yuri Gershtein and Yurii
Maravin for their guidance and help, and the staffs at Fermilab and collaborating institutions.
\end{acknowledgments}


\end{document}